# ARTIFICIAL INTELLIGENCE IN THE GLOBAL SOUTH (AI4D): POTENTIAL AND RISKS

P. J. Wall, ADAPT Centre, Trinity College Dublin, Ireland, pj.wall@tcd.ie

Deepak Saxena - BITS Pilani, India, deepak.saxena@pilani.bits-pilani.ac.in

Suzana Brown- SUNY Korea, South Korea, suzana.brown@sunykorea.ac.kr

**Abstract:** Artificial intelligence is becoming more widely available in all parts of the world. This has created many previously unforeseen possibilities for addressing the challenges outlined in the Sustainable Development Goals in the Global South. However, the use of AI in such contexts brings with it a unique set of risks and challenges. Among these are the potential for Governments to use such technologies to suppress their own people, and the ethical questions arising from implementing AI primarily designed and developed in the Global North into vastly different social, cultural, and political environments in the Global South. This paper examines the key issues and questions arising in the emerging sub-field of AI for global development (AI4D) and the potential and risks associated with using such technologies in the Global South. We propose that although there are many risks associated with the use of AI, the potential benefits are enough to warrant detailed research and investigation of the most appropriate and effective ways to design, develop, implement, and use such technologies in the Global South. We conclude by calling for the wider ICT4D community to continue to conduct detailed research and investigation of all aspects of AI4D.

## 1. INTRODUCTION

The term Artificial Intelligence (AI) is generally accepted to have been first used at a conference in 1956[1]. Over the years there have been many definitions of AI, with perhaps the most widely accepted being by Rich (1985) who posited AI as the study of how to make computers do things which, at the moment, people are better. AI typically uses pattern recognition, reasoning, and decision-making under complex conditions, and often deals with noisy data, uncertainties, ill-defined problems, and the need for in-situ solutions (Venkatasubramanian, 2019). AI has been used in many fields over the last three decades including the natural and physical sciences, computer science, engineering, natural language processing, and medicine. It is widely accepted that AI is advancing rapidly, and there is currently a great deal of interest in the commercial potential of AI. The impact of AI will undoubtedly be significant and highly disruptive in many industries including finance, healthcare, manufacturing, retail, supply chain, logistics, and agriculture.

AI is also likely to have a significant impact on the Global South, providing new and unforeseen opportunities to address many of the problems in healthcare, education, and social services common in such resource-constrained environments. The aspiration in many low-income countries is to leverage AI in order to achieve transformation by changing underlying systems of development and towards inclusion by addressing both the symptoms and causes of inequality. These technologies are bound to have the potential to empower individuals and communities, leading to greater social change, an improved quality of life, and strengthened public health and education systems.

---

[1] https://courses.cs.washington.edu/courses/csep590/06au/projects/history-ai.pdf

However, the increased use of AI in the Global South is also likely to bring with it many potential risks and challenges. Among these risks are that Governments will use such technologies inappropriately to monitor and suppress their own populations, perhaps even going so far as to use AI-based weapons for conflict and war. In addition, ethical questions are bound to arise as AI designed and developed in the Global North is expected to be used in hugely different social, cultural and political contexts in the Global South. This is likely to raise concerns around the potential to exacerbate digital inequalities and further embed notions of digital colonialism in the Global South (Kwet, 2019).

The main purpose of this paper is to initiate a discussion about the potential and risks of AI in the Global South. The objective is to present a broad overview of the topic which will provide a starting point for future discussion and research on what is bound to become a key sub-field in ICT4D over the coming years. The paper proceeds as follows: Firstly, we discuss the potential of AI and associated technologies in the Global South. This is followed by consideration of the risks associated with such technologies in this context. We conclude with our thoughts on the current state of the AI4D sub-field, and we make a call for additional research in order to expand our knowledge on this important topic.

## 2. THE POTENTIAL OF AI IN THE GLOBAL SOUTH

The increasing use and potential of AI in the Global South has been predicted for many years (e.g. Wall et al., 2013; Zheng et al., 2018), with a variety of emerging potentialities for AI and other advanced technologies which should be embraced (Walsham 2012). AI may be envisaged as a system's ability to interpret data, to build learning models from such data, and subsequently use these models to achieve specific objectives, incorporating flexible adaptation (Kaplan & Haenlein, 2019). AI does influence a large part of modern life in the Global North, having daily impact on who we interact with on social networks, the movies and TV we watch, the music we listen to, and the route we take when driving to work or school. Such AI does not work in isolation; they collect data using other advanced technologies and complement others with their decision-making capabilities. Beyond technology, the use of AI is influenced by human actors and policy/regulatory infrastructure. In that sense, AI may be seen as complex socio-technical assemblage (Cantrell & Zhang, 2018)

One of the most controversial uses of AI is in unmanned aerial vehicle (UAV) or drones. Applications of drones in the Global South are mainly restricted to delivery vehicles in agriculture and healthcare. However, drones are being increasingly used as weapons and tools for monitoring and surveillance in both international and domestic law enforcement. However, drones can also be used for wildlife preservation and scientific research. Another potentially AI-related advanced technology gaining use in the Global South is the Internet of Things (IoT). The IoT is an interconnection between the physical object and the digital world. Kevin Ashton defined IoT in 1999 as a set of ordinary objects combined with the sensors and Radio Frequency Identification Technology (RFID), which provides an infrastructure to uniquely identify and link physical objects to their virtual representations on the Internet[2]. The IoT is an integral part of the future Internet ecosystem which is likely to have a huge impact on healthcare and education in the Global South over the coming years.

Perhaps the most important potential of AI-enabled technology in the Global South is in the area of smart or precision agriculture for increasing crop productivity (Ampatzidis et al., 2020). This is especially important for countries where a large portion of people earn their living from agriculture. Drones can be used for aerial observation, sensing, and the spraying of pesticides, and various IoT sensors can provide real-time data on farms that enable farmers to make informed decisions regarding farm inputs usage. The data thus collected can be uploaded to the cloud for further analysis regarding the suitability of a crop in a particular region at a particular time of the year. Drones can also be used for surveying and mapping forests and biodiversity and can therefore help in conservation, as well as the accurate prediction of potential fires or floods based on the analysis of past data. For instance, Dutta et al., (2016) analyse data from various sources and provide a highly accurate estimates on bush-fire incidence in Australia. Such estimates can be used in planning mitigation and early evacuation, with the potential of this technology being particularly relevant in the Global South.

Also of relevance is the way AI is being used in intelligent energy management. IoTs can help in managing the entire energy lifecycle including power generation, power transmission, power distribution, and demand side management (Zhou et al., 2016). The data collected from various IoT devices (e.g., power grid equipment, GIS, weather sensors) can be uploaded to the cloud where power generation companies can run complex simulations and forecast power requirements. Data collected from IoT sensors may also be used for fault detection, electric device health monitoring, predictive maintenance, and power quality monitoring (Zheng et al., 2018).

Widespread adoption of these AI-based technologies will be facilitated by the roll out of fibre and 4G networks which is currently taking place across much of the Global South. This, combined with cheap and powerful smartphones built specifically for the African market, raises the possibility that AI has the potential to become more ubiquitous in such resource-constrained settings. Over the coming years advances in this technology will mean that AI will be able to do more, with this being particularly true in the field of healthcare where remote diagnosis of various medical conditions and diseases, faster and more reliable transfer of larger amounts of data, more sophisticated monitoring and control of data and the ability to conduct a variety of advanced medical scans with a mobile phone becoming a likelihood. An example of this is the acceleration sensors inbuilt within the phone which will enable patients and health workers to interact more closely, and mobile apps with the potential to produce and manage considerable amounts of data by using the camera and various other measuring and sensing devices to automate the logging of personal health states (Benferdia & Zakaria 2014). It is also suggested that future possibilities for AI and healthcare in the Global South include remote diagnosis and crowd sourcing for health (Latif, Rana et al. 2017), more big, open, and real-time data, the use of field sensors and embedded computing, more social media, more crowd-sourcing models, and 3D printing (Heeks 2014).

---

[2] https://www.rfidjournal.com/that-internet-of-things-thing

# 3.         RISKS OF AI IN THE GLOBAL SOUTH

Despite the huge potential of AI in the Global South, there exists a less pleasant future agenda which Zheng (2018) refers to as the "dark side" of ICT4D. This involves AI being used for surveillance and control, which includes big data, social media, and cloud computing. Such applications of the technology will further enhance the capacity of authorities and commercial entities to access a wide variety of personal data. Zheng (2018) makes the claim that there has been little discussion on this important issue to date, and on the implications of AI4D which may be controlled more frequently by capital and commercial interests in future. In addition, there is the possibility that such technology will be used by governments to enforce their ideologies and maintain control of populations in a variety of ways. This is particularly relevant as the application of AI may have a dramatic effect on many countries in the Global South (Zheng et al., 2018).

Furthermore, although AI is potentially beneficial, it is currently expensive to develop and implement. This cost of implementing such advanced technologies is the biggest risk for resource-constrained countries in the Global South. While AI has the potential to help in precision agriculture and forest preservations, the initial infrastructure cost is huge when compared to the cost of more traditional methods. This is where the role of technology companies and transnational organizations comes into play; such organizations may need to provide some basic infrastructure to support low-income countries in the short run to create a market in the longer run.

Another example of potential risk is the implementation of AI-based sensors, robots and other devices on farms that may cause harm to farm animals and external wildlife (Ryan, 2019). These devices could upset, injure, or even kill livestock and local wildlife. Many of the devices also have the potential to emit toxic material, fumes, and waste into their surrounding environment. An additional concern is that AI algorithms may exclude the land external to the farm and therefore cause adverse effects to the general environment (Antle, et al., 2015) such as an encroachment on habitats or general pollution. Therefore, the ecological and social effects of AI implementation in agriculture are significant (Kosier 2017) and should not be ignored.

In addition, many AI algorithms are opaque and often beyond human understanding (Hagras, 2018; Mackenzie, 2019). Undifferentiated use of such AI, particularly AI which has been trained on data from countries in the Global South, may result in the loss of opportunities (e.g., rejection of a mortgage or a parole) for some in the Global South. Moreover, drones have the potential to cause injuries and damage to people and properties if used incorrectly or unethically. This is partly due to the absence of a legal and governance framework across the Global South. Even in the Global North, where there are some guidelines (e.g., EU guidelines on trustworthy AI), these risks persist.

# 4.         DISCUSSION AND CONCLUSIONS

The transformative potential of AI and other advanced technologies, combined with the existing critical mass of technological infrastructure, might inspire some to be optimistic about the prospects for AI in the Global South. However, many are pessimistic as it has previously proven to be difficult to implement, sustain and scale any type of technology for global development projects. Indeed, Heeks (2018, p. 103) goes further by claiming that "*most ICT4D projects fail*". Furthermore, many such projects prove to have either limited or unsustainable impacts on development (Chipidza & Leidner 2019). Similar issues are bound to apply to implementing, using, sustaining, and scaling AI in the Global South.

This raises the obvious question as to how it might be possible to successfully design, develop, implement, and use AI and other advanced technologies in the Global South. Considering AI as assemblage, one way to achieve this may be through the adoption of a transdisciplinary or multidisciplinary perspectives in AI4D research (Walsham, 2017, Zheng, et al., 2018). Such an approach involves welcoming other disciplines with open arms (Walsham, 2012) and with respect (Walsham, 2017). This is important because it will expand the ICT4D field of study into many non-traditional settings (Walsham, 2012). According to both Zheng and Walsham, such linkages should be developed between ICT4D and the fields of computer science, development studies, ethics, anthropology, human geography, development economics and rural development. These links are needed as the nature of the sub-field of AI4D and the technologies being used are dynamically evolving, with social media, AI and the IoT (Zheng et al., 2018), as well as mobile sensing devices to automate the logging of personal health states (Benferdia & Zakaria, 2014), crowdsourcing for health (Heeks, 2014; Latif, Rana et al., 2017), increasing amounts of big data, open data and real-time data becoming more common in the Global South.

Also of relevance to this discussion is the claim that there has been a significant amount of reinvention of the wheel in ICT4D research (Zheng, et al. 2018), with new entrants to the field tending to neglect earlier research simply because technologies have changed rapidly. This is a mistake which should not be repeated in our work with AI4D. Taking stock of ICT4D research and capitalising on existing knowledge would appear to be vital and may also enable the AI4D field to move forward quicker without repeating earlier pitfalls (Zheng, et al., 2018).

Based on what we have presented in this paper we conclude that although there are many risks associated with the use of AI and other advanced technologies in the Global South, the potential benefits of these technologies are enough to warrant us to be cautiously optimistic. However, we recommend proceeding with caution so as not to repeat previous mistakes which have resulted in most ICT4D projects failing or underachieving. We are hopeful that this will be achieved by adopting the transdisciplinary and multidisciplinary perspectives as espoused by Geoff Walsham and others.

We conclude this short paper by making a call for the IFIP WG 9.4 community, as well as the wider ICT4D community, to continue to conduct detailed research and investigation of all aspects of AI4D. In particular, the emphasis should focus on a study of the most appropriate and effective ways to design, develop, implement, and use AI and other advanced technologies in the Global South, and the broader ethical implications associated with the use of such advanced technologies in this context.

## ACKNOWLEDGEMENTS

This research was partially conducted at the ADAPT SFI Research Centre at Trinity College Dublin. The ADAPT SFI Centre for Digital Content Technology is funded by Science Foundation Ireland through the SFI Research Centres Programme and is co-funded under the European Regional Development Fund (ERDF) through Grant #13/RC/2106_P2.